\documentclass[10pt,journal,compsoc]{IEEEtran}
\usepackage{amsmath,amsfonts,amssymb,mathrsfs,bm,}
\usepackage{algorithmic,enumerate}
\usepackage{algorithm}
\usepackage{array}
\usepackage{textcomp}
\usepackage{url,stfloats}
\usepackage{color}
\usepackage{graphicx}
\usepackage{cite}
\usepackage{subfigure}
\newtheorem{theorem}{Theorem}

\newtheorem{remark}{Remark}
\newtheorem{definition}{Definition}
\newtheorem{example}{Example}

\newtheorem{assumption}{Assumption}
\newtheorem{problem}{Problem}

\begin{document}

\title{Inverse learning of black-box aggregator for robust Nash equilibrium}

\author{Guanpu Chen, Gehui Xu, Fengxiang He, 
Dacheng Tao, 
Thomas Parisini,
Karl Henrik Johansson
\thanks{G. Chen 
, G. Xu, 
and Karl H. Johansson are with Division of Decision and Control Systems, 
School of	Electrical Engineering and Computer Science, KTH Royal Institute of Technology, and also with Digital Futures, Stockholm 100 44, Sweden. (e-mail: guanpu@kth.se,
 gehui@kth.se, 
 kallej@kth.se)}
\thanks{F. He is with Artificial Intelligence and its Applications Institute, 
School of Informatics, University of Edinburgh, Edinburgh EH8 9AB, Scotland.	(e-mail: F.He@ed.ac.uk)}
\thanks{D. Tao is with the College of Computing \& Data Science, Nanyang Technological University, Singapore 639798.
(e-mail: dacheng.tao@ntu.edu.sg)}
\thanks{T. Parisini is with Department of Electrical and Electronic Engineering,
	Imperial College London, London SW7 2AZ, UK, and also with Department
	of Engineering and Architecture, University of Trieste, Trieste 34127, Italy. (e-mail: t.parisini@imperial.ac.uk)}
 
 \vspace{-1.2 em}
}

%

\maketitle

\begin{abstract}

In this note, we investigate the robustness of Nash equilibria (NE) in multi-player aggregative games with coupling constraints.
There are many algorithms for computing an NE of an aggregative game given a known aggregator. When the coupling parameters are affected by uncertainty, robust NE need to be computed.
We consider a scenario where players' weight in the aggregator is unknown, making the aggregator kind of ``a black box''. 
We pursue a suitable learning approach to estimate the unknown aggregator
by proposing an inverse variational inequality-based relationship.
We then utilize the counterpart to reconstruct the game and obtain first-order conditions for robust NE in the worst case.
Furthermore, we characterize the generalization property of the learning methodology via an upper bound on the violation probability.
Simulation experiments show the effectiveness of the proposed inverse learning approach.

\end{abstract}


\section{Introduction}

Multi-player game-theoretical models have gained popularity as they offer a comprehensive understanding of interactions in multi-agent systems. 
Aggregative games play a particularly important role \cite{paccagnan2016distributed,lei2020distributed,9721063} in non-cooperative games, where each player’s payoff is dependent on both its action and an
aggregator of all players' weighted actions.
Indeed, more and more aggregative games enjoy widespread applications, such as demand
response management \cite{ye2016game},  congestion communication control \cite{barrera2014dynamic}, and public environmental investigation \cite{cornes2016aggregative}. In this connection, many algorithms have been developed and deployed to compute a Nash equilibrium (NE) in aggregative games within several different scenarios \cite{koshal2016distributed,liang2017distributed,fabiani2020robustness,belgioioso2020distributed,9784880}.

In actual application contexts, though, uncertainty inevitably emerges, for example,  in electric vehicle charging \cite{yang2015noncooperative}, security resource allocation \cite{nikoofal2012robust}, or moving target defense \cite{cheng2022single}. 
Therefore, the robustness of solutions is not an option but is a necessity. 
Robust game theory\cite{aghassi2006robust}, where uncertainty arises in players' payoffs or strategies,  draws inspiration from robust optimization \cite{bertsimas2011theory,calafiore2006scenario}.

More specifically, robust equilibrium seeking in multi-player games can be divided into two main categories. 
One consists in enforcing satisfaction of all uncertain feasibility \cite{chen2021distributed,xu2023algorithm,10178136}. This viewpoint originates from the deterministic robust optimization \cite{bertsimas2011theory}, in order to reveal the worst-case solution subject to all possible conditions. 
Usually, such approaches employ robust counterparts to reconstruct the problem against uncertainty. 
The other category is to address uncertainty with a high probability   \cite{fele2020probably,fabiani2022probabilistic,pantazis2023priori}, stemming from scenario programming \cite{calafiore2005uncertain,calafiore2006scenario}, to
randomly extract finite samples from uncertain feasibility and reconstruct the problem under their intersection.  
This viewpoint usually concerns how to find the supported samples.



Nonetheless, consider a practical scenario where 
a well-developed algorithm like \cite{koshal2016distributed,liang2017distributed,fabiani2020robustness,belgioioso2020distributed,9784880} has been already deployed to compute NE given an aggregative game. The system automatically runs after deployment, and the algorithm returns an NE corresponding to the given parameters. 
In this setting, players' weight in the aggregator turns out to be unnecessarily known to the public thus making the aggregator kind of ``a black box". When parameters in the system are affected by uncertainty, the equilibrium-seeking algorithm still returns an NE given a perturbed parameter, but robustness is lost. The internal knowledge of the game model is indispensable to achieve robustness. However, the black-box aggregator prevents us from directly using the existing methods \cite{chen2021distributed,xu2023algorithm,10178136,fele2020probably,fabiani2022probabilistic,pantazis2023priori} to address uncertainty. A suitable learning approach to ``disassemble" the black-box aggregator and estimate players' weight therein is thus needed to obtain robust NE.

In this note, we focus on approaching robust NE in a class of aggregative games with uncertainty. We first formulate the game model under uncertainty and give the concept of robust NE in the worst case. Then, we consider the situation where the players' weight in the aggregator is unknown to the public. We propose to address the robustness  by recovering the black-box aggregator from data  and reconstructing the problem from a robustness perspective.  

The main contribution is threefold.




\begin{itemize}
	\item A learning method is proposed to estimate players' weight in the black-box aggregator. By assembling {perturbed} parameters and  computed NE into data, we obtain an inverse variational inequality relationship {(Theorem \ref{t1})}. We employ a slack variable as the loss, the minimization of which enables to state an inverse optimization problem.

	\item Through the counterpart, the robustness of NE is addressed 
 by transforming the recovered aggregative game with uncertainty into a deterministic worst-case model {(Theorem \ref{t2})}. We show first-order conditions of robust NE, making gradient-based approaches usable.
 
 
 \item To characterize the learning performance, a generalization guarantee  of the proposed method is provided by using the violation probability. 
A generalization bound indicates not only the independence of the uncertainty distribution, but also the exponential convergence as the dataset size increases {(Theorem \ref{t3})}.
\end{itemize}

The note is organized as follows.
Section~II gives the problem formulation whereas Section~III illustrates our learning approach. Section~IV addresses the robustness and Section~V presents the generalization aspects. Section~VI shows the effectiveness of our methodology via extensive numerical results and Section~VII gives a few concluding remarks.
\section{Problem Formulation}\label{formulation}
In this section, we show the game model with uncertainty, the robustness of NE, and the problem statement.


\subsection{Game Model with Uncertainty}

{Consider a multi-player aggregative game $\mathscr  G$, where players are indexed by $\mathcal I=\{1,\dots, N\}$. For  player $i\in\mathcal I$, its strategy is given by the variable $x_i\in\mathbb R^n$ and the others' strategies are collected in $\bm x_{-i}\in\mathbb R^{(N-1)\times n}$. Let $\bm x\in\mathbb R^{Nn}$ stand for all players' strategies.
Player $i$ has a 
payoff function $f_i:\mathbb R^{Nn}\rightarrow \mathbb R$.
The map of an aggregator $\sigma: \mathbb R^{Nn}\rightarrow \mathbb R^n$ is defined by
\begin{align}
\sigma(\bm x)=\sum\limits_{i=1}^N \beta_i x_i,
\end{align}
where $\beta_i\in\mathbb{R}$   corresponds to player $i$'s weight in the aggregator. Take  
$\bm \beta=(\beta_1,\dots,\beta_N)\in\mathbb{R}^N$. 
For $i\in\mathcal I$, let $J_i:\mathbb R^n\times\mathbb R^n\rightarrow \mathbb R$ be a continuously differentiable function and suppose that $J_i(x_i,\sigma(\bm x))$ is convex in $x_i$.
In an aggregative game $\mathscr G$, player $i$'s payoff function satisfies 
$f_i(x_i,\bm x_{-i})=J_i(x_i,\sigma(\bm x))$.

We introduce the parameter $\bm \alpha=(\alpha_1,\dots,\alpha_N)$  with {$\alpha_i\in\mathbb{R}^n$}. Given $\bm x_{-i}$, the constraint set for $x_i$ is defined by 
\begin{align*}
\Omega_{i,\bm \alpha}(\bm x_{-i})=\{x_i\in\mathbb R^{n}_+:~ \alpha_i^Tx_i\leq b-\!\sum_{j\neq i, j=1}^N\!\!\! \alpha_j^Tx_j\},
\end{align*} 
where $b$ is a scalar parameter. Take $\bm{\mathrm{A}}=\prod_{i=1}^N\mathrm{A}_i\subseteq \mathbb {R}^{Nn}$ to represent the uncertainty in parameter $\bm\alpha$. Then, 
given others' strategies $\bm x_{-i}$, player $i$ solves the following problem:
\begin{equation}
\begin{aligned}\label{origin}
    \min_{x_i}\quad &
    J_i(x_i,\sigma(\bm x))\\
    \mathrm {s.t.}\quad & x_i\in\Omega_{i, \bm \alpha}(\bm x_{-i}),~{\bm \alpha\in\bm {\mathrm{A}}.}
\end{aligned}
\end{equation}

The overall coupling constraint can be denoted by\
$$
\Omega_{\bm\alpha}=\{\bm x\in\mathbb R^{Nn}_+: ~\sum_{i=1}^N\alpha_i^T x_i\leq b\},
$$
which means that players' strategies are subject to
 a coupling constraint as resource allocation \cite{liang2017distributed,belgioioso2020distributed}.
The uncertainty in \eqref{origin} indicates that the linear inequality in the coupling constraint $\Omega_{\bm \alpha}$ should be satisfied for all  $\alpha_i\in {\mathrm{A}_i}$, $i\in\mathcal I$. Concretely, we investigate
a typical uncertain feasibility
$$\mathrm{A}_i=\{\alpha_i\in\mathbb R^{ n}: D_i\alpha_i\leq d_i\},\quad i\in\mathcal I,$$   
where $D_i\in\mathbb R^{m_i\times n}$ is a matrix equipped with normalized rows and $d_i\in\mathbb R^{m_i}$ is a vector. In fact, $\mathrm{A}_i$ is a polyhedron enclosed by hyperplanes and the dimension $m_i$ reflects the number of hyperplanes.

\begin{assumption}\label{ass}
The constraint set $\Omega_{\bm \alpha}$ has a nonempty interior point under all the uncertainty $\bm \alpha\in\bm{\mathrm{A}}$.
\end{assumption}
}
Notice that if Slater's condition holds for all uncertain feasibility $\bm{\mathrm{A}}$, it is also true with a fixed parameter $\bm\alpha$.


\subsection{Robustness of Nash Equilibrium}  

Given a fixed parameter $\bm \alpha\in\bm{\mathrm {A}}$, minimizing the payoff subject to the coupling constraint $x_i\in\Omega_{i,\bm \alpha}(\bm x_{-i})$ turns out to be a deterministic problem.  It is already widely studied in the past decade \cite{chen2021distributed,liang2017distributed,ye2016game}.  
We take $\mathscr G_{\bm \alpha}$ as the deterministic aggregative game under a fixed
 $\bm \alpha$.
We first revisit the well-known definition of the generalized Nash equilibrium (GNE) in $\mathscr G_{\bm \alpha}$
\cite{facchinei2010generalized}.

\begin{definition}\label{d1}
	A strategy profile $\bm x_{\bm \alpha}^*$ is a GNE of the aggregative game $\mathscr G_{\bm \alpha}$ if, for all $i\in\mathcal I$, we have
	\begin{align*}
 f_i(x^*_{\bm \alpha,i},\bm x^*_{\bm \alpha,-i})\leq f_i(x_i,\bm x^*_{\bm \alpha,-i}),~\forall  x_i\in\Omega_{i,\bm \alpha}(\bm x_{\bm \alpha,-i}^*).
 \end{align*}
\end{definition}
Definition \ref{d1} indicates that $\bm x_{\bm \alpha}^*$ is a GNE of  $\mathscr G_{\bm \alpha}$ if 
no player can get a better
 payoff by modifying its strategy unilaterally.
{Then, the pseudo-gradient can be given by
\begin{align}\label{pseudo}
\bm F(\bm x)&=\begin{pmatrix}F_1(\bm x)\\
\vdots\\
F_N(\bm x)
\end{pmatrix}=
\begin{pmatrix}
\nabla_{x_1}f_1(x_1,\bm x_{-1})\\
\vdots\\
\nabla_{x_N}f_N(x_N,\bm x_{-N})
\end{pmatrix}\nonumber\\
&=\begin{pmatrix}
\nabla_{x_1}J_1(\cdot,\sigma)+\beta_1\nabla_{\sigma}J_1(x_1,\cdot)\\
\vdots\\
\nabla_{x_N}J_N(\cdot,\sigma)+\beta_N\nabla_{\sigma}J_N(x_N,\cdot)
\end{pmatrix}.
\end{align}
With the pseudo-gradient information, we can establish the connection between  GNE and the first-order condition in $\mathscr G_{\bm\alpha}$. 
As we know, there exist various ways to seek a GNE.
One of the most accepted approaches is to seek a variational GNE (vGNE), which requires a unified multiplier when deriving the Nash Karush-Kuhn-Tucker (KKT) condition \cite{chen2021distributed,facchinei2010generalized,liang2017distributed}.  
Technically, the definition of a vGNE can be found as follows.
\begin{definition}
    A strategy profile $\bm x^*_{\bm \alpha}$ is a vGNE of $\mathscr G_{\bm \alpha}$ if and only if there exists {$\lambda^*\in\mathbb{R}$} such that, for all $i\in\mathcal I$,
\begin{align*}
\bm 0_n\in &  \nabla_{x_i}J_i(\cdot,\sigma(\bm x^*_{\bm \alpha}))\!+\!\beta_i\nabla_{\sigma}J_i(x^*_{\bm \alpha,i},\cdot)\!+\!\lambda^*\alpha_i\!+\!\mathcal N_{\mathbb R_+^n}(x^*_{\bm \alpha,i}),\\
0\geq & \sum_{i=1}^N\alpha_i^Tx^*_{\bm \alpha,i}-b\perp \lambda^*.
\end{align*}
\end{definition}
It follows from
\cite[Theorem 4.8]{facchinei2010generalized} that with the convexity of payoffs $J_i$ and Assumption 1,
a vGNE $\bm x^*_{\bm \alpha}$  of $\mathscr G_{\bm \alpha}$ is equivalent to a solution to a variational inequality (VI) problem VI$(\Omega_{\bm\alpha}, \bm F)$, \textit{i.e.}, to find a vector $\bm x^*_{\bm \alpha}\in\Omega_{\bm \alpha}$ such that
\begin{equation}\label{VI}
\bm F(\bm x^*_{\bm \alpha})^T(\bm x-\bm x^*_{\bm \alpha})\geq 0,~\forall \bm x\in\Omega_{\bm\alpha}.
\end{equation}
In this view, many works  are devoted to the algorithm design for seeking vGNE  with coupling constraints \cite{ye2016game,liang2017distributed,chen2021distributed,xu2023algorithm}. 
Here we do not restrict 
the monotonicity of $\bm F$ since we mainly focus on the equivalence between GNE and $\operatorname{VI}$ solutions. Details for existence and uniqueness can be found in \cite{facchinei2010generalized,facchinei2007finite}.



Now recall the uncertainty in  game $\mathscr G$. Clearly, different {perturbed} parameters $\bm\alpha$ from uncertain feasibility $\bm{\mathrm {A}}$ yield different vGNE $\bm x^*_{\bm \alpha}$ of a deterministic game $\mathscr G_{\bm \alpha}$. 
Therefore, game $\mathscr G$ in \eqref{origin} should be solved under all uncertain feasibility $\bm \alpha\in\bm{\mathrm {A}}$, and we need a solution in the worst case.
We introduce the following concept for the robust GNE.


\begin{definition}
	A strategy profile $\bm x^*$ is a robust  GNE (rGNE) of the aggregative game $\mathscr G$ if, for all $i\in\mathcal I$, we have
	$$f_i(x^*_i,\bm x^*_{-i})\leq f_i(x_i,\bm x^*_{-i}),~\forall  x_i\in\Omega_{i,\bm \alpha}(\bm x_{-i}^*), ~\forall \bm\alpha\in\bm{\mathrm {A}}.$$
\end{definition}

It is important to find an rGNE since it gives an acceptable solution for all players against the worst case \cite{fabiani2020robustness}. 
There have been several works devoted to robust NE seeking in multi-player games. One viewpoint consists in enforcing satisfaction of all uncertainty \cite{chen2021distributed,xu2023algorithm,10178136}. It originates from deterministic robust optimization \cite{bertsimas2011theory} to reveal the worst-case solution subject to all uncertainty. 
Another way is to satisfy the uncertainty with a high probability, stemming from scenario programming \cite{calafiore2005uncertain,calafiore2006scenario}. 
Randomly extract finite
samples from uncertain feasibility, and reconstruct the problem under their intersection \cite{fele2020probably,fabiani2022probabilistic,pantazis2023priori}. This viewpoint usually concerns
how to find the supported samples.

\subsection{Problem Statement} 
The main goal of this paper is to compute an rGNE of the aggregative game $\mathscr G$ in \eqref{origin} against uncertainty.
Since robustness is required, it is significant to logically reconstruct the
problem with the internal knowledge of the game model. However, conditions may not be  always perfect in reality and we consider the following practical scenario.

Given a fixed parameter $\bm \alpha$, suppose that a well-developed algorithm like \cite{koshal2016distributed,liang2017distributed,fabiani2020robustness,belgioioso2020distributed,9784880} has been already deployed to compute a vGNE of $\mathscr G_{\bm\alpha}$. The system automatically runs after deployment and the algorithm returns a vGNE with the given parameter. 
Such an input-output process, from the given parameter $\bm\alpha$ to the corresponding vGNE $\bm x^*_{\bm \alpha}$, yields that an outsider does not need the internal knowledge of the system. As a result, some structures of the game model may not be accessible to an outsider. Here we focus on that players' weight $\bm \beta$ in the aggregator $\sigma(\bm x)=\sum_{i=1}^N\beta_ix_i$ becomes unknown to the public, which makes the aggregator $\sigma$ a black box. 

When the parameter $\bm\alpha$ suffers uncertainty in $\bm{\mathrm {A}}$, the vGNE-seeking algorithm still returns a vGNE according to the given condition but will lose robustness. What an outsider indeed needs is an rGNE, serving as a worst-case solution under all feasibility $\bm{\mathrm {A}}$.
As mentioned, the knowledge of the game model is indispensable to achieve robustness, since one needs the structure knowledge to reconstruct the problem against uncertainty. However, the black-box aggregator prevents us from directly using the existing methods \cite{chen2021distributed,xu2023algorithm,10178136,fele2020probably,fabiani2022probabilistic,pantazis2023priori} to finish the job.

Hence, there should be a learning approach to disassemble the black-box aggregator $\sigma$ and recover players' weight $\bm\beta$ before investigating the robustness. Recall what we have: {perturbed} parameters $\bm\alpha$ from uncertainty $\bm{\mathrm {A}}$ as inputs and corresponding vGNE $\bm x^*_{\bm \alpha}$ computed by the deployed solver as outputs. 
They constitute a data point $(\bm \alpha,\bm x^*_{\bm \alpha})$. On this basis, a data-driven approach is required to estimate the black-box part before achieving robustness.

{The problem to solve in this paper can be stated as follows.
\begin{problem}\label{p1}
    Given data points $(\bm \alpha,\bm x^*_{\bm \alpha})$ composed of the {perturbed} parameters and the computed vGNE, develop a method to learn the black-box aggregator $\sigma$ and obtain an rGNE $\bm x^*$ of game $\mathscr G$ in \eqref{origin} under uncertainty.
\end{problem}
}

In terms of Problem \ref{p1}, we will address the following three concerns in the sequel: i) to propose a novel learning method based on an inverse VI-based relationship; ii) to solve the robustness with 
respect to the worst-case situation; iii) to measure the generalization of our learning method.

\section{Inverse Learning}\label{learning}
In this section, we provide an inverse VI-based learning approach to reveal players' weight $\bm\beta$ in the black-box aggregator $\sigma(\bm x)$. Recall the data $(\bm\alpha,\bm x_{\bm\alpha}^*)$ composed of a set of parameters $\bm\alpha\in\bm{\mathrm {A}}$ and corresponding vGNE $\bm x_{\bm\alpha}^*$.
In fact, the learning task is to recover the mapping from $\bm\alpha$ to $\bm x_{\bm\alpha}^*$.

\subsection{Inverse VI-based Relationship}

Consider the VI-based relationship in \eqref{VI}. Given any fixed parameter $\bm\alpha\in\bm{\mathrm {A}}$, a vGNE $\bm x_{\bm\alpha}^*$ of the deterministic game $\mathscr G_{\bm\alpha}$ serves as a solution to  VI$(\Omega_{\bm\alpha}, \bm F)$, that is,
$
\bm F(\bm x^*_{\bm \alpha})^T(\bm x-\bm x^*_{\bm \alpha})\geq 0,~\forall \bm x\in\Omega_{\bm\alpha}.
$
The following theorem indicates an inverse relation which will help construct the learning model.

\begin{theorem}\label{t1}
	Consider the deterministic game  $\mathscr G_{\bm \alpha}$ under a fixed parameter $\bm \alpha$. Under Assumption \ref{ass}, $\bm x^*_{\bm \alpha}$ is a vGNE of $\mathscr G_{\bm \alpha}$  if and only if there exists a scalar $ \gamma\leq 0$ such that
%
\begin{equation}
\begin{aligned}\label{relation}
\bm F( \bm x^*_{\bm \alpha})^T\bm x^*_{\bm \alpha}- \gamma b&\leq 0,\\
F_i(\bm x^*_{\bm \alpha})-\gamma \alpha_i &\geq \bm 0_n,~\forall i\in\mathcal I.
\end{aligned}
\end{equation}
\end{theorem}	
\noindent\textbf{Proof. }
It follows from the expression in \eqref{VI} that
\begin{align}\label{VI1}
\bm F(\bm x^*_{\bm \alpha})^T\bm x^*_{\bm \alpha}\leq \bm F(\bm x^*_{\bm \alpha})^T\bm x,~\forall \bm x\in\Omega_{\bm\alpha},
\end{align}
where we notice that $\bm x^*_{\bm \alpha}$ is a constant.
\eqref{VI1} should be satisfied for all   $\bm x\in\Omega_{\bm\alpha}$, which means the inequality holds when the right-hand side takes on the minimum.
\begin{align}\label{VI2}
\bm F(\bm x^*_{\bm \alpha})^T\bm x^*_{\bm \alpha}\leq \min_{\bm x\in\Omega_{\bm\alpha}} \bm F(\bm x^*_{\bm \alpha})^T\bm x.
\end{align}
Hence, a new optimization problem  arises:
\begin{equation}\label{newopti}
\min_{\bm x\in\Omega_{\bm\alpha}} \bm F(\bm x^*_{\bm \alpha})^T\bm x.
\end{equation}
Then, we investigate \eqref{newopti} from a duality perspective. Recalling the coupling constraint set $\Omega_{\bm\alpha}=\{\bm x\in\mathbb R^{Nn}_+: ~\sum_{i=1}^N\alpha_i^T x_i\leq b\}$, the Lagrangian function can be designed as follows:
\begin{align*}
\mathcal L_1(\bm x,\gamma,\bm \mu)&=\bm F(\bm x^*_{\bm \alpha})^T\bm x-\gamma(\sum_{i=1}^N \alpha_i^Tx_i- b)+\sum_{i=1}^N\mu_i^T x_i\\
&=\sum_{i=1}^N(F_i(\bm x^*_{\bm \alpha})^T-\gamma\alpha_i^T+\mu_i^T)x_i +\gamma b,
\end{align*}
where the multipliers  $0\geq \gamma\in\mathbb R$
and $\bm 0_{nN}\geq\bm \mu=col\{\mu_1,\dots,\mu_N\}\in\mathbb R^{nN} $
 are employed for the inequality constraints $\sum_{i=1}^N \alpha_i^T x_i\leq b$ and the non-negative orthant $\bm x\in\mathbb R^{Nn}_+$, respectively.

Under Assumption \ref{ass},  
given
any fixed $\bm\alpha$, the dual gap of $\mathcal L_1$ vanishes. Followed by the duality relation, the minimum of $ \bm F(\bm x^*_{\bm \alpha})^T\bm x$ subject to $\bm x\in\Omega_{\bm\alpha}$ equals to
\begin{align*}
\max_{\gamma,\bm \mu\leq 0} ~\gamma b\quad
s.t. ~F_i(\bm x^*_{\bm \alpha})-\gamma\alpha_i+\mu_i=\bm 0_n,~\forall i\in\mathcal I.
\end{align*}
In fact, we can remove the multiplier $\bm \mu$ and thus simplify the expression of the above optimization as 
\begin{align}\label{new}
\max_{\gamma\leq 0} ~\gamma b\quad
s.t. ~F_i(\bm x^*_{\bm \alpha})-\gamma\alpha_i\geq \bm 0_n,~\forall i\in\mathcal I.
\end{align}

{So far, we derived  the above dual problem \eqref{new} corresponding to the optimization $\min_{\bm x\in\Omega_{\bm\alpha}} \bm F(\bm x^*_{\bm \alpha})^T\bm x$, and there is no duality gap due to Assumption \ref{ass}.
}
Hence, the inequality in \eqref{VI1}, which has been equivalently transferred to the inequality in \eqref{VI2}, can be further rewritten as
\begin{align}\label{VI3}
\bm F(\bm x^*_{\bm \alpha})^T\bm x^*_{\bm \alpha}\leq \max_{\gamma\in\Gamma_{\bm\alpha}} \gamma b,
\end{align}
where $
 \Gamma_{\bm\alpha}=\{\gamma\leq 0:~F_i(\bm x^*_{\bm \alpha})-\gamma\alpha_i\geq \bm 0_n,~\forall i\in\mathcal I\}.$
Again, if the inequality in \eqref{VI3} needs to be satisfied  with  the right-hand side taking on the maximum, then we should merely ensure that there exists at least one feasible $\gamma\in\Gamma_{\bm\alpha}$. Hence, the inequality in \eqref{VI3} finally becomes 
\begin{equation}\label{finally}
    \bm F(\bm x^*_{\bm \alpha})^T\bm x^*_{\bm \alpha}\leq\gamma b.
\end{equation}
We combine the requirement \eqref{finally} and
$\Gamma_{\bm\alpha}$, and obtain that if $\bm x^*_{\bm\alpha}$ is a vGNE of game $\mathscr G_{\bm\alpha}$ satisfying the relation in \eqref{VI1}, then there exists $\gamma\leq 0$ leading to the consequence.

The reverse result can be proven similarly using weak duality properties since all the aforementioned conversions are equivalently conducted. 
\hfill$\square$

 

{Most vGNE-seeking solvers compute a vGNE {asymptotically} 
since there is rarely a closed-form solution in complicated multi-player games. A numerical vGNE may not  satisfy an exact solution, considering the computational errors and other biases. Thus, we relax (\ref{VI})} with  a slack coefficient $\delta\geq 0$ such that 
$
    \bm F(\bm x^*_{\bm \alpha})^T(\bm x-\bm x^*_{\bm \alpha})+\delta\geq 0,\forall \bm x\in\Omega_{\bm\alpha}.
$
Accordingly, the relation \eqref{relation} in Theorem \ref{t1} can be correspondingly revised. There exists $\gamma\leq 0$ such that
\begin{equation}
\begin{aligned}\label{relax-relation}
\bm F(\bm x^*_{\bm \alpha})^T\bm x^*_{\bm \alpha}- \gamma b&\leq \delta,\\
F_i(\bm x^*_{\bm \alpha})-\gamma \alpha_i &\geq 0,~\forall i\in\mathcal I.
\end{aligned}
\end{equation}
Similar to the proof in Theorem \ref{t1}, the modified {relation in} \eqref{relax-relation} can also be rigorously guaranteed. 
Here, if the mapping from $\bm\alpha$ to $\bm x_{\bm\alpha}^*$ is recovered well enough, then given $\bm\alpha$, the prediction should also be exactly $\bm x_{\bm\alpha}^*$. Hence, $\delta$ also stands for a role of loss, which means the more accurate the mapping, the smaller the loss between the prediction and the practical equilibrium point. 

\subsection{Data-driven Optimization}
 
We show how to use the relation in \eqref{relax-relation} to design a learning approach for the unknown weight $\bm\beta$ in the black-box aggregator. We recall what knowledge we already have: 1) a data point composed of perturbed $\bm \alpha$ and the computed vGNE $\bm x^*_{\bm \alpha}$; 2) a relaxed relation with loss in (\ref{relax-relation}).
By the expressions of $\bm \beta$ in 
 the pseudo-gradient $\bm F$ in \eqref{pseudo}, clearly, $\bm F$ belongs to a parametric family indexed by $\bm\beta$.
Thus, we can rewrite $\bm F(\bm x)$ as $\bm F(\bm x;\bm \beta)$ to describe this dependence, and naturally suppose that $\bm F(\bm x;\bm \beta)$ is continuous in $\bm \beta$.


On this basis, we design an inverse VI-based optimization.
Here, the data point is $(\bm \alpha,\bm x^*_{\bm \alpha})$, variables are the unknown weight $\bm\beta$, the auxiliary variable $\gamma$, and the slack variable $\delta$ as the loss. Hence,
\begin{equation}
\begin{aligned}\label{inverse_opt}
\min_{\bm \beta,\gamma,\delta} \quad&|\delta|\\
\mathrm{s.t.}~ &\bm F(\bm x^*_{\bm \alpha};\bm \beta)^T\bm x^*_{\bm \alpha}- \gamma b\leq \delta, ~\gamma\leq 0,\\
&F_i(\bm x^*_{\bm \alpha};\bm \beta)-\gamma \alpha_i \geq 0,~\forall i\in\mathcal I.
\end{aligned}
\end{equation}

{We turn back to robust considerations for the uncertain feasibility $\bm{\mathrm {A}}$. Note that \eqref{inverse_opt} is derived under a fixed $\bm\alpha$, 
but the problem cannot be well-solved with only one data point $(\bm\alpha,\bm x_{\bm\alpha}^*)$. 
Fortunately, uncertainty occurs in the parameter $\bm\alpha\in\bm{\mathrm {A}}$
and provides enough data. Once the system is perturbed, there emerges a new-extracted parameter $\bm \alpha$. Then,
a well-deployed vGNE-seeking solver produces  a corresponding numerical solution $\bm x^*_{\bm \alpha}$, and  $(\bm\alpha, \bm x^*_{\bm \alpha})$ will serve as a new data point satisfying the relation  \eqref{relax-relation}. }


Therefore, we use index $k$ to represent the $k$th uncertain condition and regard  $(\bm \alpha [k], \bm x^*_{\bm \alpha[k]})$ as the $k$th data point. Also, take  $\gamma[k]$ and $\delta[k]$ as the $k$th variables to be optimized together according to \eqref{inverse_opt}. We take $\bm \delta=col\{\delta[1],\cdots\,\delta[M]\}$, $\bm \gamma=col\{\gamma[1],\cdots,\gamma[M]\}$, and $\bm\beta = \{\beta_1,\dots,\beta_N\}$ as all variables.
}
With well-defined data and variables, we can finally design a data-based learning approach:
\begin{equation}
\begin{aligned}\label{inverse_data}
\min_{\bm \beta,\bm \gamma,\bm \delta} ~&\|\bm \delta\|\\
\mathrm{s.t.}~ &\bm F(\bm x^*_{\bm \alpha[k]};\bm \beta)^T\bm x^*_{\bm \alpha[k]}- \gamma[k] b\leq \delta[k],\\
&F_i(\bm x^*_{\bm \alpha[k]};\bm \beta)-\gamma[k] \alpha_i[k] \geq 0,\quad\forall i\in\mathcal I,\\
&\gamma[k]\leq 0,~k=1,\dots,M.
\end{aligned}
\end{equation}
\begin{remark}
First, we do not request the uniqueness of vGNE $\bm x^*_{\bm \alpha}$ in  $\mathscr G_{\bm \alpha}$.
   The learning still works as long as $\bm x^*_{\bm \alpha}$ satisfies the inequality (5), even if the payoffs might yield multiple equilibria \cite{bertsimas2015data}. Second, the optimal solution to  \eqref{inverse_data} should exist, but is not necessarily unique. The tie can be broken by selecting  the one with the minimal $l_2$-norm among optimal solutions. 
    Third, the norm in the objective of \eqref{inverse_data}  is not restricted and can be determined by concrete conditions.
\end{remark}
\begin{remark}
If the necessary convexity can be maintained, the learning approach is capable of being extended for nonlinear aggregators in $\bm x$, that is, $\sigma(\bm x)=\sum_{i=1}^N\beta_i g_i(x_i)$. Such a form of an aggregator actually still maintains explicit parametric properties in variable $\bm\beta$. As for more general cases, for example $\sigma(\bm x)=\sum_{i=1}^N g_i(x_i)$, the learning approach \eqref{inverse_data} based on parametric estimation may fail. Some non-parametric learning approaches like kernel methods or neural networks would help in learning $g_i$.  

\end{remark}

We provide a typical case study  for interpretation.

\begin{example}
Consider an aggregative game with $N$ electricity users in the demand of energy consumption problem \cite{ye2016game,liang2017distributed}. User $i$ adopts {$x_i\in \Omega_{i,\bm \alpha}(\bm x_{-i}) $}  as the energy consumption and aims to minimize its electricity cost 
$J_i(x_i,\sigma(\bm x))= l_i(x_i-m_i)^2+P(\sigma(\bm x))x_i$,
where $l_i$ and $m_i$ are constants of energy consumption, and $P= qN\sigma(\bm x) +p_0$ with 
$\sigma(\bm x)=\sum_{i=1}^N \beta_i x_i$. 
Then learning in \eqref{inverse_data} can be expressed as follows:
\begin{align}
\min_{\bm \beta,\bm \gamma,\bm \delta} ~&\|\bm \delta\|\nonumber\\
\mathrm{s.t.}~ & \sum_{i\in\mathcal I}  \big(2l_i(x^*_{i, \bm \alpha[k]}\!-\!m_i)\!+\!2qN\beta_{i}x^*_{i, \bm \alpha[k]}\!\nonumber\\
&\quad+\!qN \!\!\!\!\!\sum_{j\in\mathcal I, j\neq i}\! \!\!\beta_{j}x^*_{j, \bm \alpha[k]}\!+\!p_0\big)x^*_{i, \bm \alpha[k]}\!-\!\gamma[k] b\!\leq\!\! \delta[k],\nonumber\\
&2l_i(x^*_{i, \bm \alpha[k]}\!-\!m_i)\!+\!2qN\beta_{i}x^*_{i, \bm \alpha[k]}\nonumber\\
&\quad+\!qN \!\!\!\!\!\sum_{j\in\mathcal I, j\neq i} \!\!\!\!\beta_{j}x^*_{j, \bm \alpha[k]}\!+\!p_0
\!-\!\gamma[k] \alpha_i[k]\! \geq\! 0,~\forall i\!\in\!\mathcal I,\nonumber\\
&\gamma[k]\leq 0,~k=1,\dots,M.\nonumber
\end{align}
It is a solvable optimization problem with linear constraints.
\end{example}


\section{GNE Robustness}\label{robustcounterpart}
In this section, we address the robustness of GNE with the recovered knowledge by considering the worst-case solution.
We introduce some new notations after learning.  Take 
$\hat\sigma(\bm x)=\sum_{i=1}^N \hat \beta_i x_i\in\mathbb R^n$ for the aggregator,
where $\hat{\bm \beta}=\{\hat \beta_1,\dots,\hat\beta_N\}$ is revealed by  \eqref{inverse_data}.  Accordingly,  take $\widehat{\mathscr G}$  to represent the game model with uncertainty after learning, \textit{i.e.}, each player $i$ has to solve the following problem: 

\begin{align}\label{robust}
    \min_{x_i}~J_i(x_i,\hat\sigma(\bm x))~
    \mathrm{s.t.} ~ x_i\in\Omega_{i, \bm \alpha}(\bm x_{-i}),~{\bm \alpha\in\bm{\mathrm {A}}.}
\end{align}

{Then, we seek an rGNE of game $\widehat{\mathscr G}$ in the worst case, that is, the robustness of Nash equilibrium in \eqref{robust} satisfying all possible uncertainties in the feasibility $\bm{\mathrm {A}}$.} In this view, we consider transforming the uncertain problem \eqref{robust} into a deterministic model. 
{By the virtue of deterministic robust optimization,}
the following theorem shows how to construct a robust counterpart.
\begin{theorem}\label{t2}
Under Assumption 1, a strategy profile $\bm x^*$ is an rGNE of game $\widehat{\mathscr G}$ \eqref{robust} if and  only if there exists $\bm y^*=col\{y_1^*,\dots,y_N^*\}$ with $y_i^*\in\mathbb R^{m_i}$ such that $(\bm x^*,\bm y^*)$ is a GNE of the following deterministic game:
\begin{equation}
\begin{aligned}\label{robust_counter}
\min_{x_i\in \mathbb{R}^n_+, y_i\in \mathbb{R}_+^{m_{i}}} & J_i(x_i,\hat \sigma(\bm x))\\
\mathrm{s.t.}\quad&\sum_{i=1}^N y_i^Td_i \leq b,~D_i^Ty_i\!-\!x_i=\bm 0_{n},~ \forall i\in\mathcal I.
\end{aligned}
\end{equation}
\end{theorem}
\noindent\textbf{Proof. }
Recall the expression of the coupling constraint $\Omega_{\bm\alpha}$ under all uncertain feasibility $\bm\alpha\in\bm{\mathrm {A}}$. If all possible constraints hold in \eqref{robust}, then it can be equivalently regarded as
\begin{align}\label{worstcase}
\max_{\bm\alpha\in\bm{\mathrm {A}}}\sum_{i=1}^N\alpha_i^Tx_i=\sum_{i=1}^N\max\limits_{\alpha_i\in\mathrm{A}_i}\alpha_i^Tx_i\leq b.
\end{align}
{We find a sub-optimization problem } $\max\limits_{\alpha_i\in\mathrm{A}_i}\alpha_i^Tx_i$ with the feasibility structure  $\mathrm{A}_i=\{\alpha_i\in\mathbb R^{ n}: D_i\alpha_i\leq d_i\}$.
\begin{align}\label{subopt}
\max_{\alpha_i} \alpha_i^Tx_i \quad
\mathrm{s.t.}~ D_i\alpha_i\leq d_i.
\end{align} 
Notice that here $x_i$ serves as a constant while $\alpha_i$ is variable. 
Accordingly, design a Lagrangian function with an auxiliary multiplier $y_i\in\mathbb R^{m_i}_+$.
\begin{align*}
\mathcal L_2(\alpha_i,y_i)&=-\alpha_i^Tx_i+y_i^T(D_i\alpha_i- d_i)\\
&=(y_i^TD_i-x_i^T)\alpha_i-y_i^Td_i.
\end{align*}
By Assumption \ref{ass}, the polyhedron $\mathrm{A}_i$ is nonempty and the Slater's condition is therefore satisfied in problem \eqref{subopt}. Then, the duality gap vanishes in the Lagrangian function $\mathcal L_2$. It follows from the duality theory that the maximum in \eqref{subopt} can be equivalently described by the following minimum
\begin{align}\label{dual}
\min_{y_i}~ y_i^Td_i\quad
\mathrm{s.t.}~ D_i^Ty_i-x_i=\bm 0_{n}, ~y_i\geq \bm 0_{m_{i}}.
\end{align} 
Hence, the inequality relation \eqref{worstcase} in the worst case becomes
\begin{align}\label{trans}
\sum_{i=1}^N\min_{y_i\in Y_i}~& y_i^Td_i\leq b,
\end{align} 
where the constraint $Y_i=\{y_i\in \mathbb{R}_+^{m_i}:\; D_i^Ty_i-x_i=\bm 0_{n}\}$.
It follows from \cite{bertsimas2011theory} that, the minimum on the left-hand side in \eqref{trans} can be removed. In fact, if there exists at least one qualified profile $\bm y=col\{y_1,\dots,y_N\}$, then the minimum will be naturally verified. Hence, \eqref{trans} can be rewritten as 
\begin{align}\label{trans1}
\sum_{i=1}^Ny_i^Td_i\leq b,~y_i\in Y_i,~\forall i\in\mathcal I.
\end{align}

Thus, together with the definition of set $Y_i$ for $i\in\mathcal I$, the coupling constraint in \eqref{robust} under all feasibility $\bm \alpha\in\bm{\mathrm {A}}$  can be reformulated by the following deterministic constraints:
\begin{align}
\sum _{i=1}^N y_i^Td_i\leq b,~ D_i^Ty_i-x_i=\bm 0_{n}, ~\forall i\in\mathcal I,
\end{align}
where $x_i\in \mathbb{R}^n_+$ and $y_i\in \mathbb{R}_+^{m_{i}}$. 
That is exactly \eqref{robust_counter}. 

The reverse proof can be conducted similarly when $(\bm x^*,\bm y^*)$ satisfies the deterministic formulation in \eqref{robust_counter}, because all the above procedures are equivalent transformations. \hfill$\square$


Theorem \ref{t2} shows that,
by introducing the auxiliary variable $\bm y$, the uncertain game $\hat {\mathscr G}$ can be transformed into a deterministic one in \eqref{robust_counter}. On this basis, we can obtain the first-order condition of  \eqref{robust_counter}. Technically, a strategy  $(\bm x^*, \bm y^*)$ is a GNE of \eqref{robust_counter} (an rGNE  of  $\hat{\mathscr G}$ in \eqref{robust}) if and only if there exists {$\mu^*\in\mathbb{R}_+$} and  {$\bm\omega^*=col\{\omega_i^*\}_{i=1}^N\in\mathbb{R}^{m}$}  such that, for $i\in\mathcal I$, 
\begin{equation}
\begin{aligned}\label{robust_first}
&\bm 0_n\in  \nabla_{x_i}J_i(\cdot,\hat{\sigma}(\bm x^*))\!+\!\hat{\beta}_i\nabla_{\hat{\sigma}}J_i(x^*_{i},\cdot)\!-\!\omega_i^*\!+\!\mathcal N_{\mathbb R_+^n}(x^*_{i}),\\
&\bm 0_{m_i}\in   \mu^*d_i+D_i\omega_i^*+\mathcal N_{\mathbb R_+^{m_i}}(y^*_{i}),\\
&0\geq  \sum_{i=1}^Nd_i^Ty^*_{i}-b\perp \mu^*,\\
&\bm 0_{n}=D_i^Ty^*_{i}-x_i^*.
\end{aligned}
\end{equation}

Now we can say that, after learning the black-box aggregator and deriving the robust counterpart, we are finally  able to seek an rGNE of the original game $\mathscr G$ in \eqref{origin} via the above first-order condition. There is no need for a detailed derivation of the solver design since such designs have been already given in many developed works \cite{koshal2016distributed,liang2017distributed,fabiani2020robustness,belgioioso2020distributed,9784880}.

{For the sake of simplicity, we put together all the procedures in Algorithm 1 to compute an rGNE within a black-box aggregative game problem \eqref{origin}.}
\begin{algorithm}[htbp]
	\renewcommand{\thealgorithm}{1}
	\caption{}
	\label{a1}
	
	\textbf{Initialization}:
	aggregator $\sigma(\bm x)$, payoff functions $J_i(x_i,\sigma(\bm x))$, coupling constraint $\Omega_{\bm \alpha}$, all uncertain feasibility $\bm{\mathrm {A}}$, and a well-developed vGNE-seeking solver.
	
	\textbf{1) Data Construction}
 
\quad\textbf{for} $k=1,2,\dots,M$, \textbf{do}

\quad\quad \textit{Input}:\quad uncertain $\alpha_i[k]\in\mathrm{A}_i,~i\in\mathcal I$,

\quad\quad \textit{Solve}:\quad deterministic  game  $\mathscr G_{\bm \alpha[k]}$ by the existing  vGNE- 

\quad\quad\quad\quad\;\,\quad seeking solver,

\quad\quad \textit{Output}: vGNE $\bm x^*_{\bm \alpha[k]}$ of game $\mathscr G_{\bm \alpha[k]}$.

\quad\textbf{end for}

	\textbf{2) Inverse learning}
 
\quad \textit{Input}:\quad all data $(\bm \alpha[k],\bm x^*_{\bm \alpha[k]}),~k=1,2,\dots,M$,

\quad \textit{Solve}:\quad inverse VI-based  learning approach in \eqref{inverse_data},

    \quad \textit{Return}: estimated weight $\hat{\bm \beta}$ in black-box aggregator

	\textbf{3) Robust Counterpart}
 


\quad \textit{Solve}:\quad the deterministic game $\hat{\mathscr  G}$ in \eqref{robust_counter} with auxiliary 

\quad\quad\quad\;\,\quad variables $y_i,i\in\mathcal  I$ and estimated weight $\hat{\bm \beta}$,


\quad \textit{Return}: rGNE $\bm x^*$ of game $\mathscr G$.

\end{algorithm}

\section{Generalization Guarantee}\label{generalization}
In this section, we address the generalization capabilities of our learning approach \eqref{inverse_data}. In fact, an important aspect of a learning method is its ability to generalize to real data, that is, whether its estimation can perform well on new data. 

\subsection{Violation Probability}

In  \eqref{inverse_data}, the generalization guarantee refers to whether the optimal solution with the used data $(\bm\alpha[k],\bm x^*_{\bm\alpha[k]})$, $k=1,2,\dots, M$, can truly represent all cases under the uncertain feasibility $\bm{\mathrm {A}}$.
In fact, no matter what sampling (referring to $\bm \alpha\in\bm{\mathrm {A}}$) is taken, it is unrealistic to fully represent the entire uncertain domain. Therefore, the optimal solution we learn by \eqref{inverse_data} may merely be applicable to those given  samplings $(\bm\alpha[k],\bm x^*_{\bm\alpha[k]})$, $k=1,2,\dots, M$,
and the optimal solution $\bm\beta$ 
 may  not be optimal under all uncertain feasibility. In other terms, there may be a new  sampled data point $(\bm\alpha[M+1],\bm x^*_{\bm\alpha[M+1]})$ for which the previously obtained optimal solution by \eqref{inverse_data} is no longer valid. 

Hence, we investigate the above fact using the following \textit{violation probability} concept \cite{calafiore2005uncertain}. For convenience, we  rewrite the constraints in problem (\ref{inverse_opt}) as follows.
\begin{align}
\mathcal X(\bm\alpha,\bm x^*_{\bm\alpha})\triangleq\{\delta,\bm\beta:~\exists \gamma,~\mathrm{s.t.}&~\bm F(\bm x^*_{\bm\alpha};\bm \beta)^T\bm x^*_{\bm\alpha}- \gamma b\leq \delta,\nonumber\\
&F_i(\bm x^*_{\bm\alpha};\bm \beta)-\gamma \alpha_i \geq 0,~\forall i\in\mathcal I\}.\nonumber
\end{align}
\begin{definition}
	Given $(\bm\beta,\delta)$, the violation probability is 
	\begin{align}
	\mathbb V(\bm\beta,\delta)\triangleq\mathbb P\{\bm \alpha\in\bm{\mathrm {A}},~(\bm\beta,\delta)\notin\mathcal X(\bm\alpha,\bm x^*_{\bm\alpha})\}.
	\end{align}
\end{definition}
Then, the generalization guarantee takes on the form: given $\epsilon\in(0,1)$, evaluate the upper bound of $\mathbb P^M(\mathbb V(\bm\beta^*,\delta^*)>\epsilon),$ 
where $(\bm\beta^*,\delta^*)$ is the optimal solution via the learning approach \eqref{inverse_data} with data
{$\{(\bm \alpha [1], \bm x^*_{\bm \alpha[1]}),\cdots, (\bm \alpha [M], \bm x^*_{\bm \alpha[M]})\}$}.
In qualitative forms, if $\epsilon$ is small enough and the upper bound in $\mathbb P$ is also suitably small, then the learning method turns out to be reliable.
If so, when we employ the inverse VI-based learning approach \eqref{inverse_data} to estimate the players' weight information in the black-box aggregator, the confidence of the learning result is high. Otherwise, the learning solution may overfit and  become too dependent on the samplings of each uncertain data point.
\begin{remark}
    Generally, one would focus on the expectation of the loss $\delta$ for the generalization error. With the definition of violation probability, we learn that $\mathbb V$  monotonically approaches zero as the value of $\delta$ increases. Besides the consensus in the monotonicity of loss, we choose to investigate $\mathbb V$ because of its representation on the solution region of $\beta$ and $\delta$, providing a more interpretable result.
\end{remark}



 \vspace {-0.3 true cm}
\subsection{Generalization Bound}

With the above statements, we consider that dataset $(\bm\alpha[k],\bm x^*_{\bm \alpha[k]})$ with $k=1,\dots,M$ is considered as random samplings extracted from the uncertain feasibility $\bm{\mathrm {A}}$ subject to a distribution $\mathbb Q$, whose exact form does not need to be known. 
We provide the following main result on the generalization bound of the inverse VI-based learning approach \eqref{inverse_data}.
\begin{theorem}\label{t3}
 Under Assumption 1, suppose that the constraint $\mathcal X(\bm\alpha,\bm x^*)$ is convex in $\bm \beta$ and the optimal solution $(\bm\beta^*,\delta^*)$ does exist. Then, for any $\epsilon\in(0,1)$, we have:
	\begin{align}\label{upper}
	\mathbb P^M(\mathbb V(\delta^*,\bm \beta^*)>\epsilon) \leq \sum_{l=0}^N \begin{pmatrix}M\\l	\end{pmatrix}\epsilon^l(1-\epsilon)^{M-l}.
	\end{align}
\end{theorem}
\noindent{\textbf{Proof. }}
By Remark 1, the objective function in \eqref{inverse_opt} or \eqref{inverse_data} is not restricted to specific norms. We take $\|\bm \delta\|_\infty$ to convert the optimization \eqref{inverse_opt} or \eqref{inverse_data} into programmings based on the computation of convex intersections:
\begin{align}\label{convex-inter}
\min_{\delta\geq 0,\bm \beta} ~\delta \quad
\mathrm {s.t.} ~ (\delta,\bm \beta)\in\bigcap_{k=1}^M\mathcal X(\bm\alpha[k],\bm x^*_{\bm \alpha[k]}).
\end{align}
We can  verify that constraint $\mathcal X(\bm\alpha,\bm x^*)$ is convex in all the variables due to the convexity in $\bm \beta$.
At this point, we have transformed the problem of obtaining the optimal parameters by data points from an uncertain set into a standard form of scenario programming \cite{calafiore2006scenario}.

On the one hand,  for random programming \eqref{convex-inter} with given samplings, the intersection of constraints is feasible and endowed with a nonempty interior point due to Slater's condition in Assumption 1.2. 
On the other hand, the uniqueness of the optimal solution is not a mandatory requirement. According to the operation in Remark 1 and \cite[Discussion 2.1.5]{campi2008exact}, the tie can be broken by selecting the one with the minimum Euclidean norm among all optimal solutions,  which can still maintain the conclusion. 

So far, the formulation  \eqref{convex-inter} is in accordance with \cite[Theorem 1]{campi2008exact}, and 
we keep up with all the same assumed conditions. 
Then, the number of sums in \eqref{upper} should originally be related to the dimension of all variables $\delta,\bm \beta$, that is, $dim(\delta,\bm \beta)$. Note that $\delta$ is a scalar and the dimension of $\bm \beta$ is the number of players in $\mathcal I$. The conclusion holds. \hfill$\square$


Theorem 3 tells that the generalization bound in \eqref{upper} is independent of the sampling distribution $\mathbb Q$ of the uncertain parameter $\bm\alpha$. Taking $\Delta=\sum_{l=0}^N \begin{pmatrix}M\\l	\end{pmatrix}\epsilon^l(1-\epsilon)^{M-l}$, we give some detailed explanations for Theorem 3 from different viewpoints.

i). \textit{Generalization perspective}: With probability $\Delta$ for sampling, the violation probability $\mathbb V$ of the learning optimal solution $(\delta^*,\bm \beta^*)$ is at most $\epsilon$. 
We can interpret this statement from ``new-data" scenarios. The conclusion answers the following question: for a fixed learning pair $(\delta^*,\bm \beta^*)$ with $M$ data, how much confidence can one hold that the pair is still the optimal solution with the least probability  $1-\epsilon$ when a new data $(\bm\alpha[M+1],\bm x^*_{_{\bm\alpha}[M+1]})$ comes? Then the answer is that one can keep the confidence at least $1-\Delta$.

ii). \textit{Data-size perspective}: A direct application of Theorem 3 is to find the smallest data size $M$ for given violation parameter $\epsilon$ and confidence threshold $\Delta$. The issue can be handled by solving the equation $\Delta=\sum_{l=0}^N \begin{pmatrix}M\\l	\end{pmatrix}\epsilon^l(1-\epsilon)^{M-l}$ and regarding $M$ as a variable. We will provide Tab.1 in Section \ref{experiment} to further illustrate the values $\Delta$ for given $\epsilon$, $M$, and $N$.
Actually, $\Delta$ corresponds to the tail probability of a binomial distribution in $M$, which exponentially converges.

{
\begin{remark}
Under the assumed convex conditions in Theorem 3, 
to assess the deviation probability $\mathbb V$, it's necessary to compute which samples are supporting under the given $M$ uncertainties. There is already mature research on how to compute support samples, and the detailed progress can be found in \cite{fabiani2022probabilistic,8299432}. Moreover, the setup in \eqref{origin} involves a well-deployed system encountering passive parameter perturbations and needs to recover the black-box part through a learning method with data. If there are potential mechanisms for acquiring more effective data, we believe that the performance of inverse learning approach \eqref{inverse_data} can be further improved.
\end{remark}
}
 
\section{Numerical Evaluation}\label{experiment}


We consider an aggregative game with $N=4$
electricity users in demand of energy consumption \cite{ye2016game,liang2017distributed}, as introduced in Example 1.  For $i=1,2,3,4$,
$$
\Omega_{i,\bm \alpha}(\bm x_{-i})=\{x_i\in\mathbb R_+:~ \alpha_i^Tx_i\leq 75-\!\!\!\!\sum_{j\neq i, j=1}^4 \alpha_j^Tx_j\}, 
$$
where the uncertain parameter $\alpha_i\in [0.1,2]$. 
User $i$ adopts {$x_i\in \Omega_{i,\bm \alpha}(\bm x_{-i}) $}  to minimize its electricity cost 
$$J_i(x_i,\sigma(\bm x))= l_i(x_i-h_i)^2+P(\sigma(\bm x))x_i,$$ where  $P(\sigma(\bm x))= qN\sigma(\bm x) +p_0$ and the system coefficients are set as $l_i=1$, $h_1=50$, $h_2=55$, $h_3=60$, $h_4=65$, $q=0.04$, and $p_0=5$. Note that the real value of players' weight in the aggregator is $\bm \beta=\{0.1,0.2,0.3,0.4\}$.

\begin{figure}[t]
	\centering	\includegraphics[width=7cm,height=4cm]{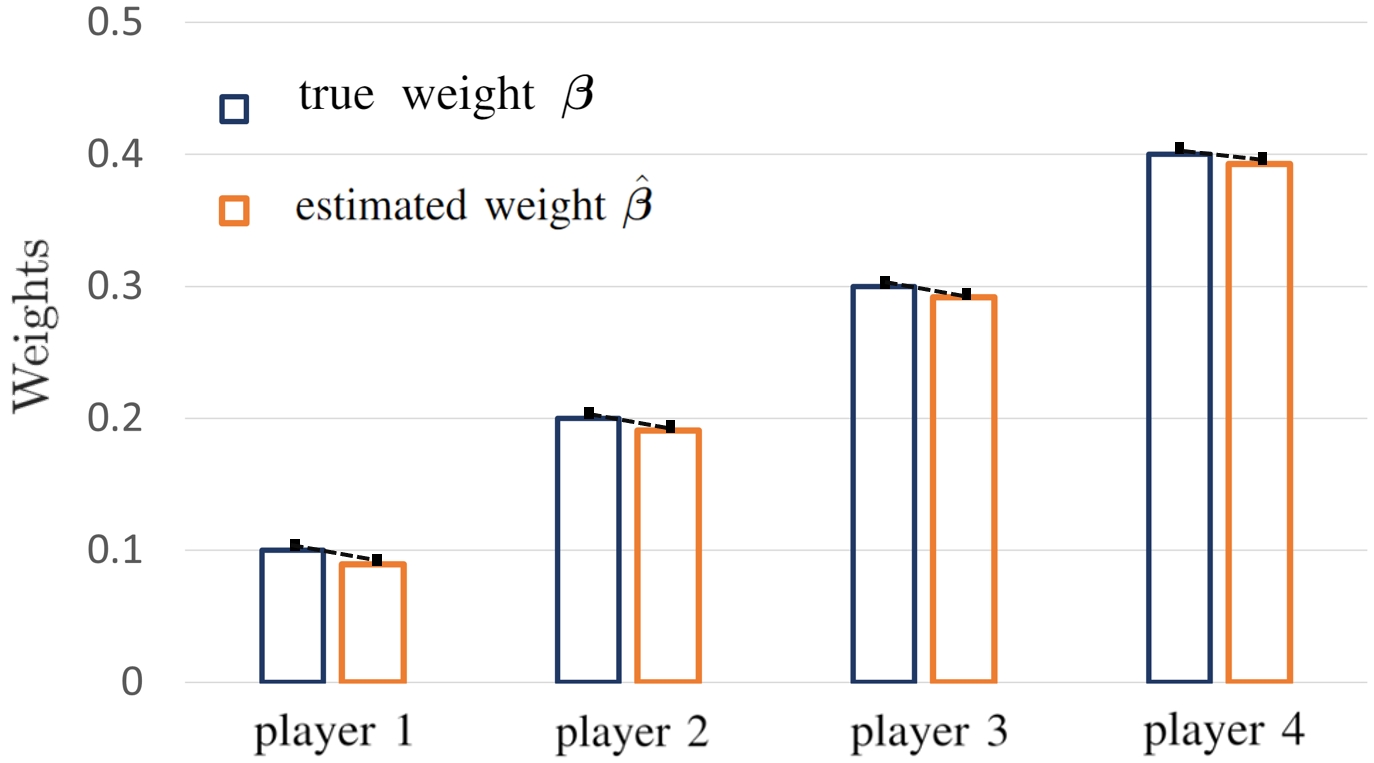}
\caption{Learning performance with data amounts $M=4$.}
\label{fig2a}
\end{figure}

\begin{figure}[t]
	\centering
	\includegraphics[width=6.5cm,height=5cm]{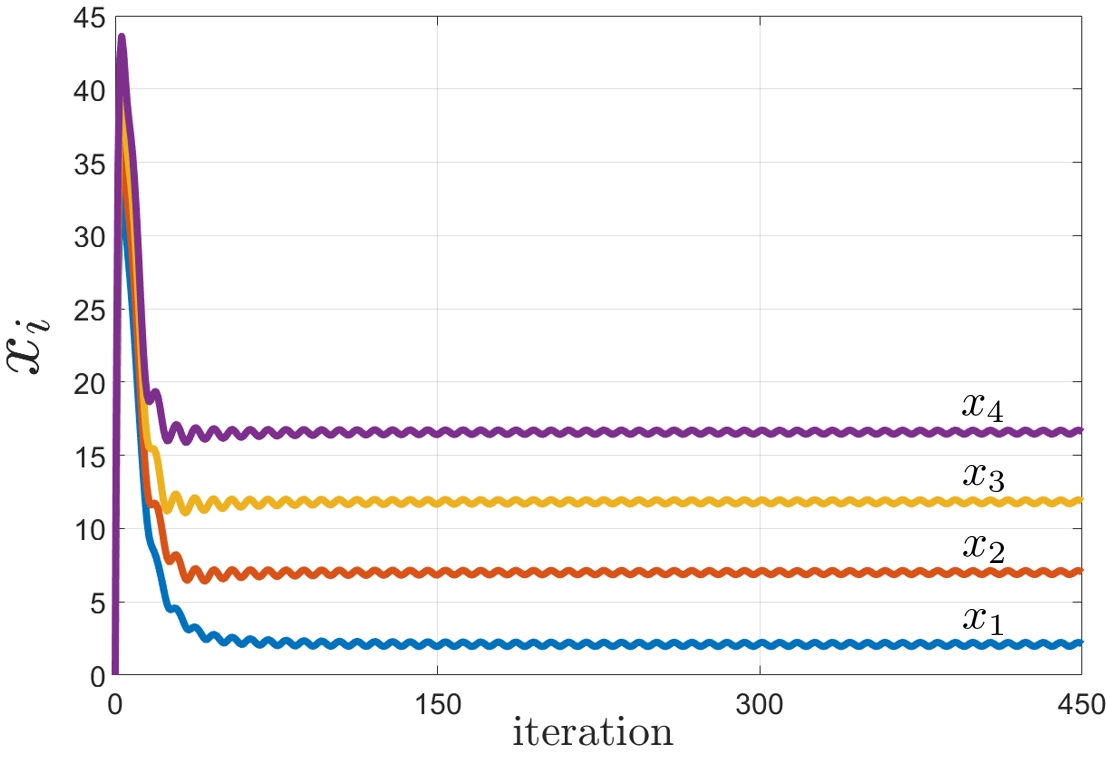}
\caption{Convergence for seeking rGNE.}
\label{fig2b}
\end{figure}

Due to the uncertain system, we first construct the dataset  $(\bm \alpha[k],\bm x^*_{\bm \alpha[k]})$ from $k=1,\dots,M$ samplings, where the equilibrium-seeking methods refer to \cite{koshal2016distributed,liang2017distributed,fabiani2020robustness,belgioioso2020distributed,9784880}. Then with all data $(\bm \alpha[k],\bm x^*_{\bm \alpha[k]})$ as input, we employ the inverse VI-based learning approach \eqref{inverse_data} to estimate players' weight $\bm\beta$ in the black-box aggregator $\sigma(\bm x)$, as similarly illustrated in Example 1.  Finally, based on learning results, we seek rGNE $\bm x^*$  by the first-order condition \eqref{robust_first}. 
We set $M=4$ and Figs.~\ref{fig2a}, \ref{fig2b} show the learning results. In Fig. \ref{fig2a}
the blue bars present the true weight $\bm \beta$ for each player while the red bars present the estimated value $\hat{\bm \beta}$ by the learning approach \eqref{inverse_data}. In detail, we get $\hat{\bm \beta}=\{0.0893,0.1907,0.2918,0.3926\}$, which shows a good learning performance close to the true value. This verifies the validity of the  VI-based inverse learning approach \eqref{inverse_data}. Afterward, we can continue to compute rGNE $\bm x^*$ with the estimated $\hat{\bm \beta}$, which is illustrated in Fig. \ref{fig2b}.
\begin{figure}[!t]
	\centering
\includegraphics[width=7.2cm,height=4cm]{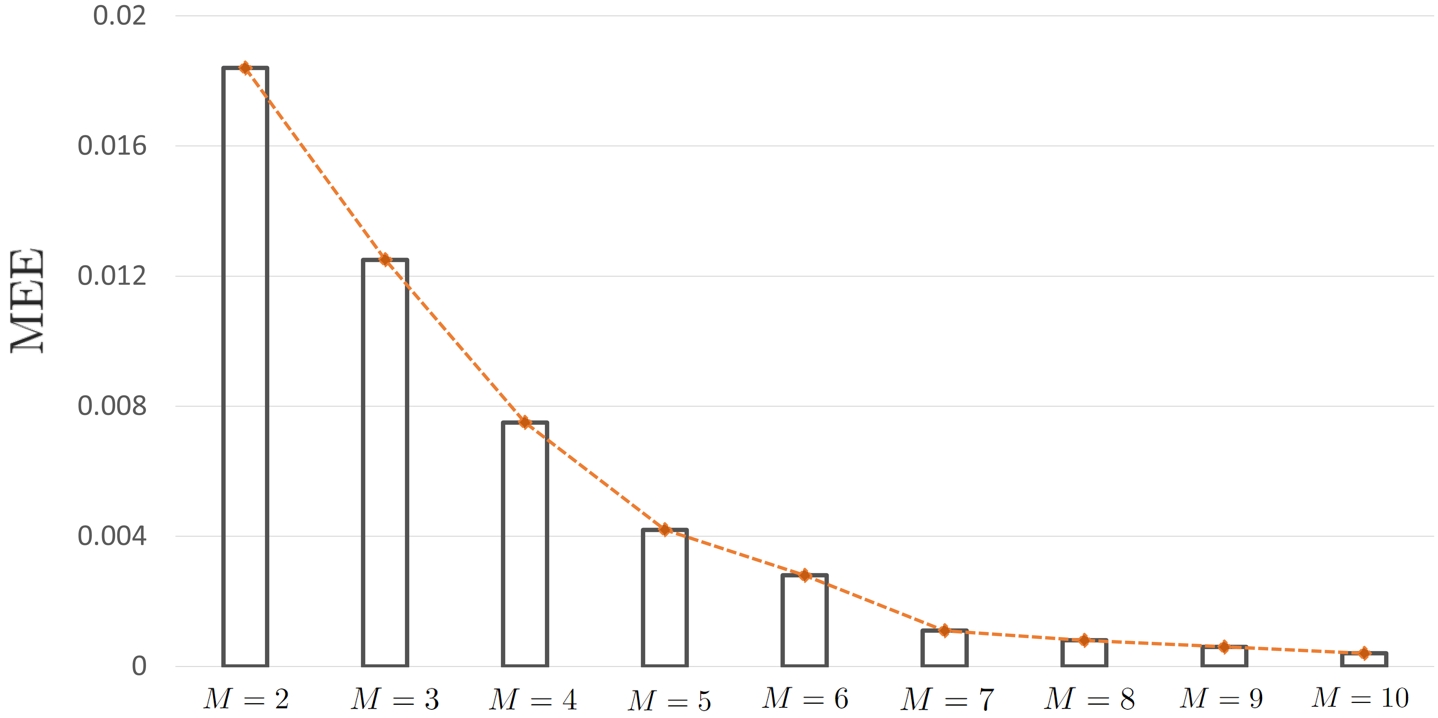}
\caption{Learning performance with different data sizes.}
\label{fig3b}
\end{figure}

We further check the performance of the learning approach \eqref{inverse_data} with different amounts of data points $(\bm \alpha[k],\bm x^*_{\bm \alpha[k]})$.
We give the following \textit{mean estimated error}
$$
  \operatorname{MEE}=\frac{1}{N} \Big({\sum_{i=1}^{N}\|\hat{ \beta_i}-{ \beta_i}\|^2}\Big)^{1/2},$$ and take $M=2,3,4,\dots,10$ to record MEE of each setting in Fig. \ref{fig3b}. The trend of MEE becomes obviously mild and close enough to zero as the amount of data increases.

 \begin{table}[htbp]
	\centering
	\normalsize
	\caption{Data size, learning error, generalization bound}
	\setlength\tabcolsep{4pt}
	\renewcommand\arraystretch{1.4}
 \scalebox{0.74}{
	\begin{tabular}{c|c|c|c|c|c}
		\hline
		\hline
		 &  $M=10$ &  $M=20$ &  $M=30$ &  $M=40$&$M=50$\\ \hline
		MEE  & 0.0004 & 0.0001 & 3.1628$*10^{-5}$ & 2.4306$*10^{-5}$ &2.4253$*10^{-5}$ \\ 
  \hline
		$\Delta$ &0.9983&0.9568&0.8245 & 0.6290&0.4312 \\
  	\hline\hline
		 &  $M=60$&$M=70$&$M=80$&$M=90$&$M=100$ \\ \hline
		MEE &2.0139$*10^{-5}$&1.8724$*10^{-5}$ &1.5397$*10^{-5}$ & 1.2256$*10^{-5}$&1.0713$*10^{-5}$\\ 
  \hline
		$\Delta$ &0.1710&0.1588&0.0880&0.0465&0.0237\\
  \hline\hline
	\end{tabular}
 }
 	\label{t123}
\end{table}
Finally, we greatly improve the numerical accuracy and provide Tab. \ref{t123} to show some numerical relation between the data amounts, learning errors MEE, and generalization bound $\Delta$. Here, take $N=4$ players, $M$ as the variant amount of data, and $\epsilon=0.1$ for violation probability. Thus, the generalization bound should be 
$\Delta=
	 \sum_{l=0}^4 \begin{pmatrix}M\\l	\end{pmatrix}{0.1}^l\cdot{0.9}^{M-l}.
  $
 We can see from Table \ref{t123} that as the dataset size increases, the value of MEE decreases, while the value of $\Delta$ goes rapidly (exponentially) to zero. This can also be regarded as a tradeoff between accuracy and confidence. These figures support the results in Theorem 3 and the associated discussions.





%
%
%
%
%
%
%
%
%

\section{Conclusions}\label{conclusion}
In this note, we proposed a novel learning scheme to seek the robust equilibrium with players' unknown weight in a black-box aggregator. We put together the data sets with two parts: {perturbed} parameters from uncertain feasibility and corresponding NE by developed solvers. We established
the learning model by an inverse variational inequality relation.
Then, we derived the robust counterpart thus obtaining the first-order conditions for
robust generalized Nashe quilibria. Also, we showed a generalization
guarantee of the proposed learning approach. The numerical results presented good performances and effectiveness of our methodology.

\bibliographystyle{IEEEtran}        
\bibliography{ref}

\begin{thebibliography}{10}
\providecommand{\url}[1]{#1}
\csname url@samestyle\endcsname
\providecommand{\newblock}{\relax}
\providecommand{\bibinfo}[2]{#2}
\providecommand{\BIBentrySTDinterwordspacing}{\spaceskip=0pt\relax}
\providecommand{\BIBentryALTinterwordstretchfactor}{4}
\providecommand{\BIBentryALTinterwordspacing}{\spaceskip=\fontdimen2\font plus
\BIBentryALTinterwordstretchfactor\fontdimen3\font minus
  \fontdimen4\font\relax}
\providecommand{\BIBforeignlanguage}[2]{{%
\expandafter\ifx\csname l@#1\endcsname\relax
\typeout{** WARNING: IEEEtran.bst: No hyphenation pattern has been}%
\typeout{** loaded for the language `#1'. Using the pattern for}%
\typeout{** the default language instead.}%
\else
\language=\csname l@#1\endcsname
\fi
#2}}
\providecommand{\BIBdecl}{\relax}
\BIBdecl

\bibitem{paccagnan2016distributed}
D.~Paccagnan, B.~Gentile, F.~Parise, M.~Kamgarpour, and J.~Lygeros,
  ``Distributed computation of generalized {N}ash equilibria in quadratic
  aggregative games with affine coupling constraints,'' in \emph{2016 IEEE 55th
  Conference on Decision and Control (CDC)}.\hskip 1em plus 0.5em minus
  0.4em\relax IEEE, 2016, pp. 6123--6128.

\bibitem{lei2020distributed}
J.~Lei, U.~V. Shanbhag, and J.~Chen, ``Distributed computation of {N}ash
  equilibria for monotone aggregative games via iterative regularization,'' in
  \emph{2020 59th IEEE Conference on Decision and Control (CDC)}.\hskip 1em
  plus 0.5em minus 0.4em\relax IEEE, 2020, pp. 2285--2290.

\bibitem{9721063}
S.~Huang, J.~Lei, and Y.~Hong, ``A linearly convergent distributed {N}ash
  equilibrium seeking algorithm for aggregative games,'' \emph{IEEE
  Transactions on Automatic Control}, vol.~68, no.~3, pp. 1753--1759, 2023.

\bibitem{ye2016game}
M.~Ye and G.~Hu, ``Game design and analysis for price-based demand response: An
  aggregate game approach,'' \emph{IEEE Transactions on Cybernetics}, vol.~47,
  no.~3, pp. 720--730, 2016.

\bibitem{barrera2014dynamic}
J.~Barrera and A.~Garcia, ``Dynamic incentives for congestion control,''
  \emph{IEEE Transactions on Automatic Control}, vol.~60, no.~2, pp. 299--310,
  2014.

\bibitem{cornes2016aggregative}
R.~Cornes, ``Aggregative environmental games,'' \emph{Environmental and
  Resource Economics}, vol.~63, no.~2, pp. 339--365, 2016.

\bibitem{koshal2016distributed}
J.~Koshal, A.~Nedi{\'c}, and U.~V. Shanbhag, ``Distributed algorithms for
  aggregative games on graphs,'' \emph{Operations Research}, vol.~64, no.~3,
  pp. 680--704, 2016.

\bibitem{liang2017distributed}
S.~Liang, P.~Yi, and Y.~Hong, ``Distributed {N}ash equilibrium seeking for
  aggregative games with coupled constraints,'' \emph{Automatica}, vol.~85, pp.
  179--185, 2017.

\bibitem{fabiani2020robustness}
F.~Fabiani, K.~Margellos, and P.~J. Goulart, ``On the robustness of equilibria
  in generalized aggregative games,'' in \emph{2020 59th IEEE Conference on
  Decision and Control (CDC)}.\hskip 1em plus 0.5em minus 0.4em\relax IEEE,
  2020, pp. 3725--3730.

\bibitem{belgioioso2020distributed}
G.~Belgioioso, A.~Nedi{\'c}, and S.~Grammatico, ``Distributed generalized
  {N}ash equilibrium seeking in aggregative games on time-varying networks,''
  \emph{IEEE Transactions on Automatic Control}, vol.~66, no.~5, pp.
  2061--2075, 2020.

\bibitem{9784880}
G.~Xu, G.~Chen, H.~Qi, and Y.~Hong, ``Efficient algorithm for approximating
  {N}ash equilibrium of distributed aggregative games,'' \emph{IEEE
  Transactions on Cybernetics}, vol.~53, no.~7, pp. 4375--4387, 2023.

\bibitem{yang2015noncooperative}
H.~Yang, X.~Xie, and A.~V. Vasilakos, ``Noncooperative and cooperative
  optimization of electric vehicle charging under demand uncertainty: A robust
  stackelberg game,'' \emph{IEEE Transactions on Vehicular Technology},
  vol.~65, no.~3, pp. 1043--1058, 2015.

\bibitem{nikoofal2012robust}
M.~E. Nikoofal and J.~Zhuang, ``Robust allocation of a defensive budget
  considering an attacker's private information,'' \emph{Risk Analysis: An
  International Journal}, vol.~32, no.~5, pp. 930--943, 2012.

\bibitem{cheng2022single}
Z.~Cheng, G.~Chen, and Y.~Hong, ``Single-leader-multiple-followers stackelberg
  security game with hypergame framework,'' \emph{IEEE Transactions on
  Information Forensics and Security}, vol.~17, pp. 954--969, 2022.

\bibitem{aghassi2006robust}
M.~Aghassi and D.~Bertsimas, ``Robust game theory,'' \emph{Mathematical
  Programming}, vol. 107, no. 1-2, pp. 231--273, 2006.

\bibitem{bertsimas2011theory}
D.~Bertsimas, D.~B. Brown, and C.~Caramanis, ``Theory and applications of
  robust optimization,'' \emph{SIAM Review}, vol.~53, no.~3, pp. 464--501,
  2011.

\bibitem{calafiore2006scenario}
G.~C. Calafiore and M.~C. Campi, ``The scenario approach to robust control
  design,'' \emph{IEEE Transactions on Automatic Control}, vol.~51, no.~5, pp.
  742--753, 2006.

\bibitem{chen2021distributed}
G.~Chen, Y.~Ming, Y.~Hong, and P.~Yi, ``Distributed algorithm for
  $\varepsilon$-generalized {N}ash equilibria with uncertain coupled
  constraints,'' \emph{Automatica}, vol. 123, p. 109313, 2021.

\bibitem{xu2023algorithm}
G.~Xu, G.~Chen, and H.~Qi, ``Algorithm design and approximation analysis on
  distributed robust game,'' \emph{Journal of Systems Science and Complexity},
  vol.~36, no.~2, pp. 480--499, 2023.

\bibitem{10178136}
M.~Fochesato, F.~Fabiani, and J.~Lygeros, ``Generalized uncertain {N}ash games:
  Reformulation and robust equilibrium seeking,'' in \emph{2023 European
  Control Conference (ECC)}, 2023, pp. 1--6.

\bibitem{fele2020probably}
F.~Fele and K.~Margellos, ``Probably approximately correct {N}ash equilibrium
  learning,'' \emph{IEEE Transactions on Automatic Control}, vol.~66, no.~9,
  pp. 4238--4245, 2020.

\bibitem{fabiani2022probabilistic}
F.~Fabiani, K.~Margellos, and P.~J. Goulart, ``Probabilistic feasibility
  guarantees for solution sets to uncertain variational inequalities,''
  \emph{Automatica}, vol. 137, p. 110120, 2022.

\bibitem{pantazis2023priori}
G.~Pantazis, F.~Fele, and K.~Margellos, ``A priori data-driven robustness
  guarantees on strategic deviations from generalised {N}ash equilibria,''
  \emph{arXiv preprint arXiv:2304.05308}, 2023.

\bibitem{calafiore2005uncertain}
G.~Calafiore and M.~C. Campi, ``Uncertain convex programs: randomized solutions
  and confidence levels,'' \emph{Mathematical Programming}, vol. 102, pp.
  25--46, 2005.

\bibitem{facchinei2010generalized}
F.~Facchinei and C.~Kanzow, ``Generalized {N}ash equilibrium problems,''
  \emph{Annals of Operations Research}, vol. 175, no.~1, pp. 177--211, 2010.

\bibitem{facchinei2007finite}
F.~Facchinei and J.-S. Pang, \emph{Finite-{D}imensional {V}ariational
  {I}nequalities and {C}omplementarity {P}roblems}.\hskip 1em plus 0.5em minus
  0.4em\relax Springer Science \& Business Media, 2007.

\bibitem{bertsimas2015data}
D.~Bertsimas, V.~Gupta, and I.~C. Paschalidis, ``Data-driven estimation in
  equilibrium using inverse optimization,'' \emph{Mathematical Programming},
  vol. 153, pp. 595--633, 2015.

\bibitem{campi2008exact}
M.~C. Campi and S.~Garatti, ``The exact feasibility of randomized solutions of
  uncertain convex programs,'' \emph{SIAM Journal on Optimization}, vol.~19,
  no.~3, pp. 1211--1230, 2008.

\bibitem{8299432}
M.~C. Campi, S.~Garatti, and F.~A. Ramponi, ``A general scenario theory for
  nonconvex optimization and decision making,'' \emph{IEEE Transactions on
  Automatic Control}, vol.~63, no.~12, pp. 4067--4078, 2018.

\end{thebibliography}

\vfill

\end{document}